\documentclass{article}
\usepackage{spconf,amsmath,graphicx}
\usepackage{hyperref}
\usepackage{subcaption}
\usepackage{color}
\usepackage{bbold}
\usepackage{dirtytalk}
\usepackage{multirow}
\usepackage{physics}
\usepackage[table]{xcolor}

\title{Extreme Audio Time Stretching using Neural Synthesis}
\twoauthors
{Leonardo Fierro, Alec Wright, Vesa V\"{a}lim\"{a}ki\sthanks{This work has been financed in part by Nokia Technologies through the DeepSlow project (Aalto University project no.~410979). L.~Fierro's work has been supported by the Aalto ELEC Doctoral School.  This work is part of the activities of the NordicSMC network (NordForsk project no.~86892). Copyright 2023 IEEE. Published in ICASSP 2023, scheduled for 4-9 June 2023 in Rhodes Island, Greece. Personal use of this material is permitted. However, permission to reprint/republish this material for advertising or promotional purposes or for creating new collective works for resale or redistribution to servers or lists, or to reuse any copyrighted component of this work in other works, must be obtained from the IEEE. Contact: Manager, Copyrights and Permissions / IEEE Service Center / 445 Hoes Lane / P.O. Box 1331 / Piscataway, NJ 08855-1331, USA. Telephone: + Intl. 908-562-3966.}}
	{Acoustics Lab, Dept. of Signal Processing and Acoustics\\
	Aalto University, Espoo, Finland}
{Matti Hämäläinen}
	{Nokia Technologies\\
	Tampere, Finland}

\begin{document}
\ninept
\maketitle
\begin{abstract}
A deep neural network solution for time-scale modification (TSM) focused on large stretching factors is proposed, targeting environmental sounds. Traditional TSM artifacts such as transient smearing, loss of presence, and phasiness are heavily accentuated and cause poor audio quality when the TSM factor is four or larger. The weakness of established TSM methods, often based on a phase vocoder structure, lies in the poor description and scaling of the transient and noise components, or nuances, of a sound. Our novel solution combines a sines-transients-noise decomposition with an independent WaveNet synthesizer to provide a better description of the noise component and an improve sound quality for large stretching factors. Results of a subjective listening test against four other TSM algorithms are reported, showing the proposed method to be often superior. The proposed method is stereo compatible and has a wide range of applications related to the slow motion of media content.
\end{abstract}
%
%
\section{Introduction}
\label{sec:intro}

Time-scale modification (TSM) refers to a change in the duration or the playback speed of a sound that does not affect its spectral characteristics, such as pitch, timbre, and brightness \cite{moulines1995non, dutilleux2011time,  driedger2016review}. If an audio signal is simply played at a different sample rate, the frequency content is deemed to be changed as the formants of the sound are moved. TSM methods are applied to avoid this phenomenon, aiming to preserve or retrieve the original spectral characteristics of the sound. 

TSM has been long used in speech, e.g.~in audio books and language--learning services \cite{verhelst1993overlap, donnellan2003speech, cohen2022speech}, music and remixing \cite{cliff2000hang}, broadcasting services \cite{dutilleux2011time}, and streaming platforms \cite{nam2018deep}. The ratio between the modified and the original time support is controlled by the TSM factor $\alpha$, defining time stretching for $\alpha>1$ and time compression for $\alpha<1$.  All of the speech TSM applications typically involve small TSM factors (0.25 $\leq\alpha\leq$ 4), and do not allow for more extreme time-scaling operations, as state-of-the-art TSM algorithms present poor audio quality at large stretch factors \cite{damskagg2017audio,roberts2021deep}. In such cases, the sounds typically present strong phasiness and heavily smeared transients, both known artifacts in phase-vocoder-based TSM implementations \cite{laroche1997phase, damskagg2017audio}. There is however an  interest for extreme audio time-stretching in applications such as slow motion \cite{moinet2013slowdio} and ambient sound generation \cite{Valimaki2018, malloy2022timbral}.

Recent work by Fierro and Välimäki \cite{fierro2022enhanced} showed that the quality of a TSM algorithm can be improved by providing a better description of its sines, transient, and noise (STN) components, which can be separated and individually processed according to their classification. It was also hinted that the potential weakness of the fuzzy phase vocoder \cite{damskagg2017audio}, which received the highest overall score both on average and for the largest tested $\alpha$ in a recent comparison of audio TSM methods \cite{roberts2021deep}, lies in the poor description and scaling of the “noisy” component of percussive events, whose smearing is not countered by the phase randomization and that would not benefit from  a full preservation as the time support of the time-stretched event would be incorrect. 

The audio synthesis landscape changed since the introduction of WaveNet, a deep generative model capable of synthesising raw audio waveforms \cite{oord2016wavenet}. The WaveNet synthesizer can be used for TSM by adjusting the number of audio samples generated per frame of the local conditioning signal, similarly to how the hop size between consecutive frames is adjusted from analysis to synthesis in traditional DSP methods. This idea was first proposed by Huang et al. \cite{huang2018timbretron}, although the TSM performance of such a model was not evaluated as time-stretching was out of its scope. 

In this work, we propose a deep learning based method for TSM capable of producing state-of-the-art results for real-world environmental sounds when large TSM factors are used. The proposed solution combines established DSP algorithms and deep learning to produce a hybrid TSM method, whose novelty lies in the neural resynthesis of the noise component, which is separated from sines and transient using the STN decomposition. 

The rest of this paper is structured as follows. Sec.~\ref{sec:STN} summarizes the STN decomposition technique. Sec.~\ref{sec:wavenet} describes the proposed WaveNet architecture, used to synthesize the stretched noise component. Sec.~\ref{sec:tsm} details the TSM pipeline. Sec.~\ref{sec:evaluation} evaluates the proposed method against four previous techniques, and Sec.~\ref{sec:conclusion} concludes.

\section{STN Decomposition}
\label{sec:STN}

The STN separation method proposed by Fierro and Välimäki \cite{fierro2022enhanced} decomposes a sound into three abstract classes: sines (tonal content), transients (impulsive events), and noise (sound nuances). The decomposition is achieved through soft spectral masks that are derived from the signal spectrogram and allow for perfect reconstruction.

To derive the class masks for an audio signal $x(n)$, median filtering is first applied to its magnitude spectrum $X(m,k)$ to highlight vertical (frequency) and horizontal (time) structures \cite{fitzgerald2010harmonic}:
\begin{multline}
X_\textrm{v}(m,k) \\ 
= \textrm{med}\Big[|X(m,k-\frac{L_\textrm{v}}{2}+1)|,...,|X(m,k+\frac{L_\textrm{v}}{2})|\Big]
\end{multline}
and
\begin{multline}
X_\textrm{h}(m,k) \\
= \textrm{med}\Big[|X(m-\frac{L_\textrm{h}}{2}+1,k)|,...,|X(m+\frac{L_\textrm{h}}{2},k)|\Big],
\end{multline}
\noindent where $\textrm{med}[\cdot]$ is the median function, and $X_\textrm{v}$ and $X_\textrm{h}$ are the resulting horizontally- and vertically-enhanced magnitude spectrograms, respectively. Parameters $L_\textrm{h}$ and $L_\textrm{v}$ are the median filter lengths (in samples) in the time and frequency directions, respectively. 

\begin{figure}[t!]
\centering
\includegraphics[width=0.8\columnwidth]{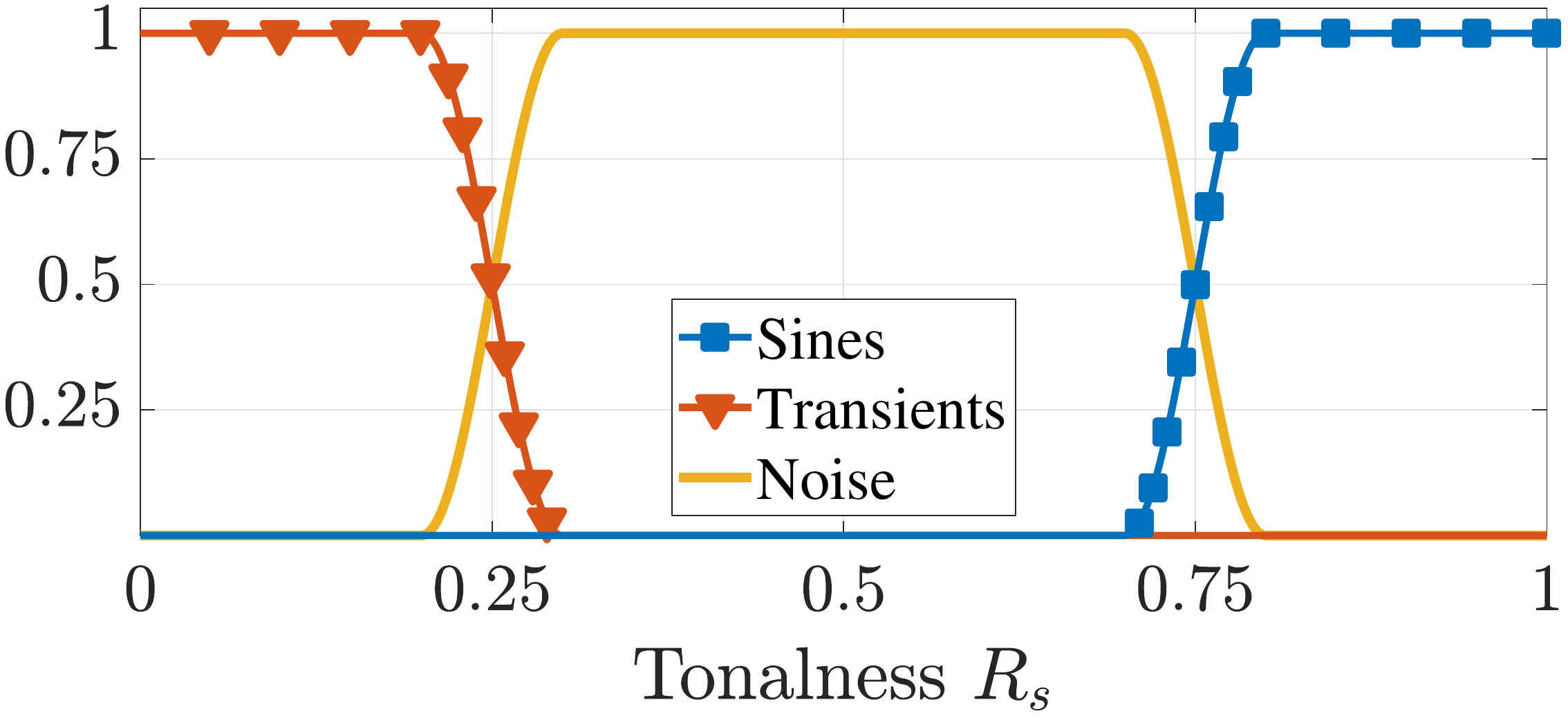}
\caption{Functions for determining soft spectral masks for STN separation, as described in \cite{fierro2022enhanced}.}
\label{fig:STNmasks}
\end{figure}

Matrices $X_\textrm{h}$ and $X_\textrm{v}$ are then used to extract the tonalness $R_\textrm{s}$ and transientness $R_\textrm{t}$ matrices with the following elements \cite{fitzgerald2010harmonic}:
\begin{equation} \label{eq:tonalness}
    R_\textrm{s}(m,k) = \frac{X_\textrm{h}(m,k)}{X_\textrm{h}(m,k)+X_\textrm{v}(m,k)}
\end{equation}
and
\begin{equation} \label{eq:transientness}
    R_\textrm{t}(m,k) = 1-R_\textrm{s}(m,k) = \frac{X_\textrm{v}(m,k)}{X_\textrm{h}(m,k)+X_\textrm{v}(m,k)},
\end{equation}

\noindent respectively. Finally, the soft masks are obtained as follows:
\begin{align}
S(m,k) = f&\left(R_\textrm{s}(m,k)\right), \\
T(m,k) = f&\left(R_\textrm{t}(m,k)\right), \\
N(m,k) = 1&-S(m,k)-T(m,k),
\end{align}
\noindent where
\begin{equation}
\begin{aligned}
f(&a) = 
\begin{cases} 
1, & \mbox{if } a \geq \beta_\textrm{U} \\
\sin^2{\Big( \dfrac{\pi}{2} \dfrac{a -\beta_\textrm{L}}{\beta_\textrm{U}-\beta_\textrm{L}}} \Big), & \mbox{if } \beta_\textrm{L} \leq a < \beta_\textrm{U} \\
0, & \mbox{otherwise},
\end{cases}
\end{aligned}
\end{equation}

\noindent which are consequently imposed onto $X(m,k)$ via element-wise multiplication to obtain the separated components. A group of functions to determine soft STN masks for $\beta_\textrm{U}$ = 0.8 and  $\beta_\textrm{L}$ = 0.7 is visualized in Fig.~\ref{fig:STNmasks}. 

\begin{figure}[t!]
\center
\begin{subfigure}[t]{0.48\columnwidth}
\centering
\includegraphics[width=\textwidth]{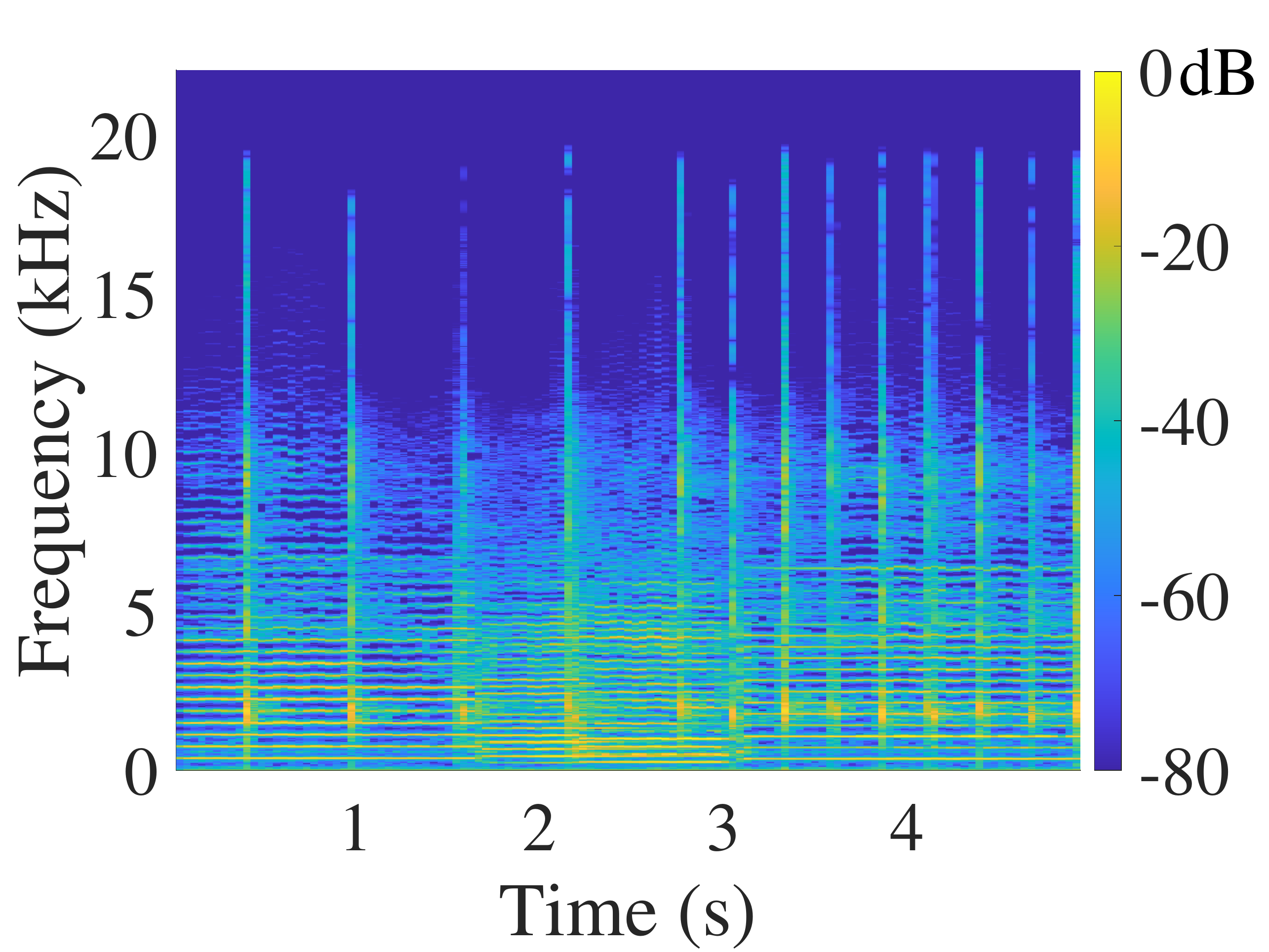}
\caption{Original}
\end{subfigure}
\begin{subfigure}[t]{0.48\columnwidth}
\centering
\includegraphics[width=\textwidth]{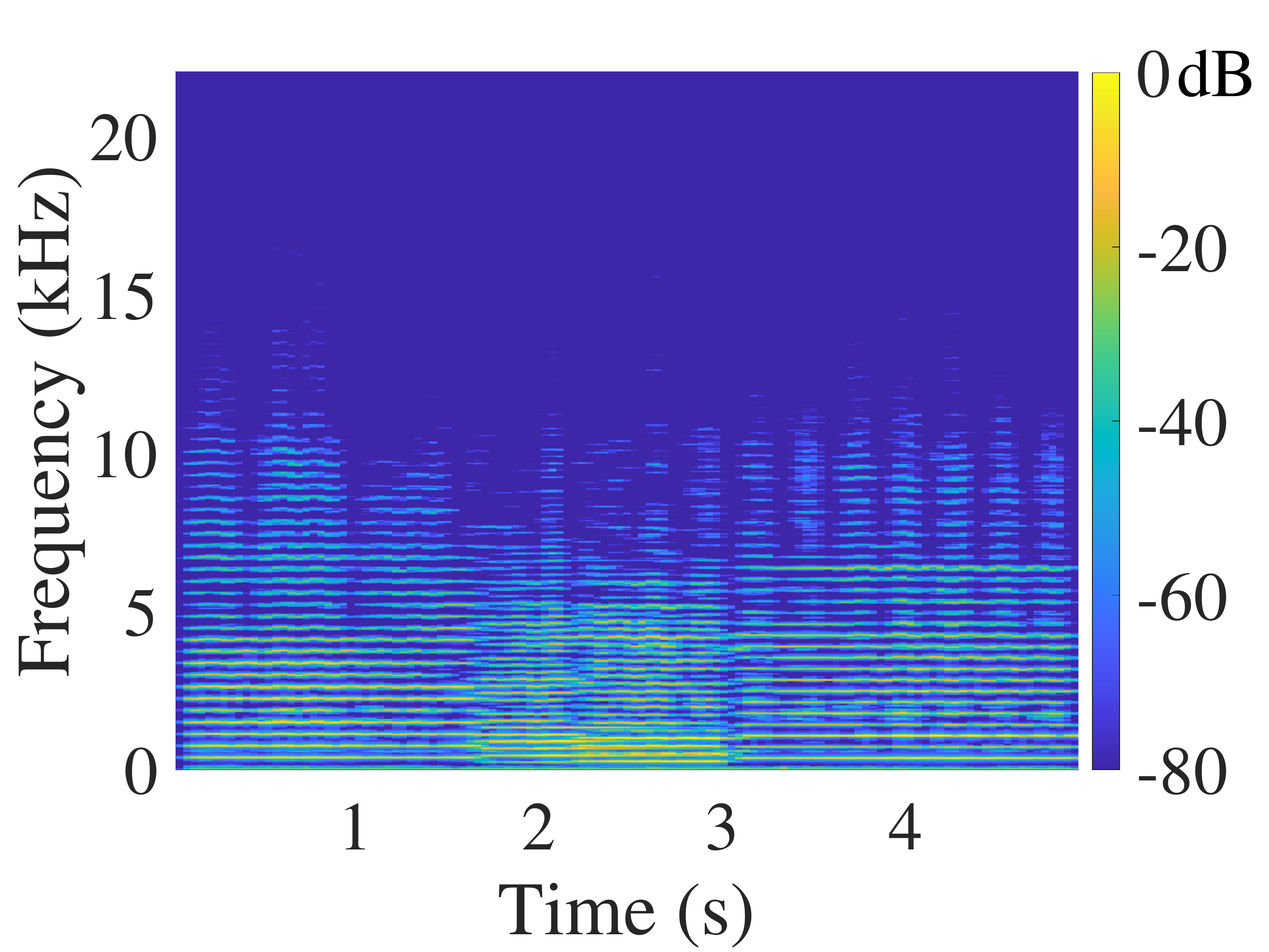}
\caption{Sines}
\end{subfigure}
\begin{subfigure}[t]{0.48\columnwidth}
\centering
\includegraphics[width=\textwidth]{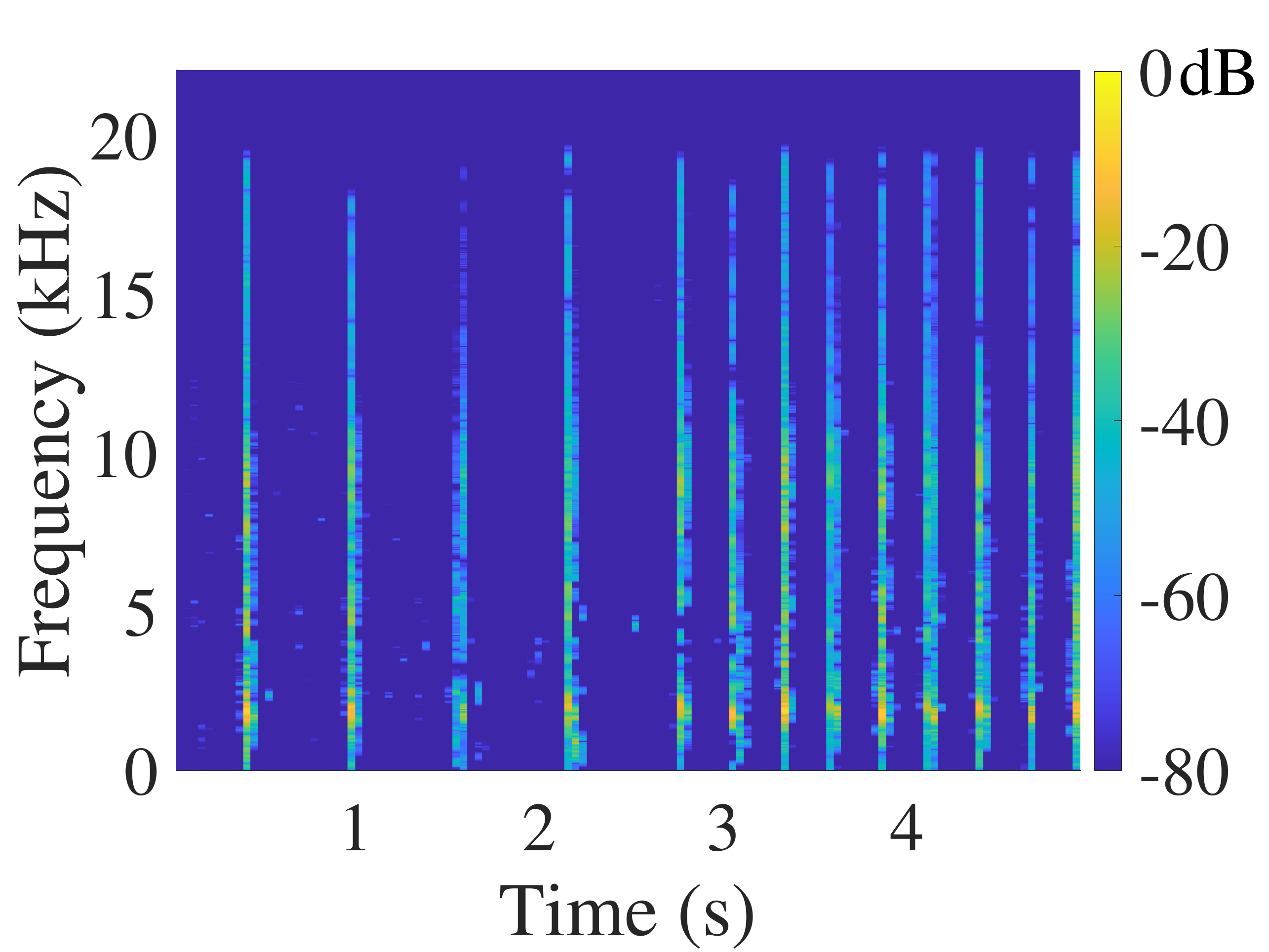}
\caption{Transients}
\end{subfigure}
\begin{subfigure}[t]{0.48\columnwidth}
\centering
\includegraphics[width=\textwidth]{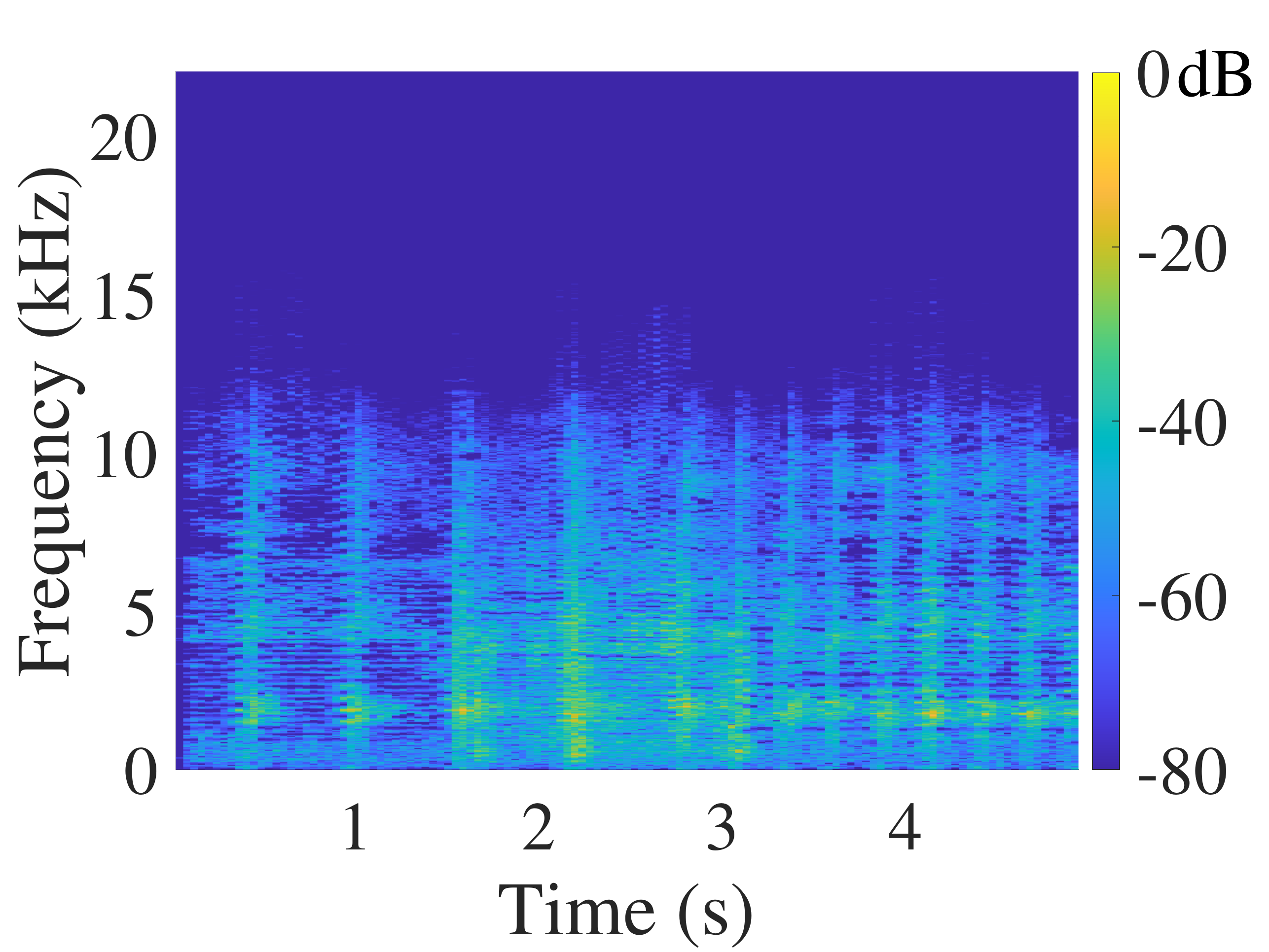}
\caption{Noise}
\end{subfigure}
\caption{\label{fig:STFTPROP} STN decomposition of a castanet and violin mixture.}
\end{figure}

This decomposition process is repeated for two consecutive stages \cite{fierro2022enhanced}. The first implements a large analysis window for better frequency resolution, separating the sines from the transient and noise residual mixture; the second uses a short analysis window for better temporal resolution, extracting the transients from the residual. An example of two-stage STN decomposition for a violin and castanet sound mixture is shown in Fig.~\ref{fig:STFTPROP}.

The STN decomposition enables the application of different TSM algorithms for each component. While traditional signal processing techniques have been optimized to deal with either the sines or the transients, the noise component remains relatively unexplored and hence is suitable for a deep learning approach. 

\section{Noise Time Stretching via Wavenet}
\label{sec:wavenet}

This section introduces a WaveNet architecture to synthesize the time-stretched noise component.

\subsection{WaveNet synthesizer}

WaveNet is an autoregressive generative model capable of synthesizing raw audio waveforms \cite{oord2016wavenet}. It was initially proposed for speech synthesis, and the model and variants thereof are still commonly used as part of audio synthesis pipelines. It consists of a stack of dilated 1D-convolutional layers, with the dilation factor increasing exponentially at each layer. The input to the network is also a raw audio waveform, with the model being trained to predict the subsequent sample in the sequence, given the prior audio samples.

Additionally, it is possible for the WaveNet model to incorporate a conditioning signal, allowing control over the audio generated by the model. There are broadly two types of conditioning information: global and local. The first describes features that influence the entirety of the generated audio, e.g.~the speaker identity in a speech synthesizer. The latter is used to represent time-variant features of the desired audio waveform, such as spectrograms and other time-frequency representations. In this work, we only consider local conditioning signals.

\subsection{WaveNet-based TSM}

Time stretching via WaveNet is achieved by adjusting the number of audio samples generated per frame of the local conditioning signal, according to $\alpha$. For example, if the conditioning signal is the spectrogram extracted from a target audio signal with a hop size of 256 samples, then the re-synthesis for $\alpha=2$ will generate 512 audio samples for each frame of conditioning signal.

In the proposed approach, the WaveNet synthesizer is used to re-synthesize the noise component of the target signal, extracted using the two-stage STN decomposition, according to the desired TSM factor $\alpha$. Three different time-frequency representations were tested as local conditioning signals for the network: spectrogram, mel-spectrogram, and Constant-Q Transform (CQT) spectrogram. Different models were trained using each of these representations, and it was found that using the CQT-spectrogram produced the highest quality results. Note that the choice of time-frequency representation is fixed and is part of the model architecture, so it cannot be changed during or after training.

The use of the CQT spectrogram as a conditioning signal for a WaveNet synthesizer was first proposed in \cite{huang2018timbretron}. The CQT transform \cite{brown1991calculation}  is a time-frequency analysis method in which the frequency bins are logarithmically spaced. Previous work has shown that the CQT is a good choice for synthesis of musical sounds \cite{huang2018timbretron} and also environmental sound classification \cite{huzaifah2017comparison}. In this work, to extract the CQT we use a 5.8 ms hop size (256 samples at the 44.1-kHz sample rate), a minimum frequency of 32.7 Hz, a maximum frequency of Nyquist (22.05\,kHz), and 48 CQT bins per octave. This results in a total of 451 frequency bins.

The network was trained at a sample rate of 44.1\,kHz, and a 10-component Mixture-of-Logistic distributions (MoL) sampling method was used to generate raw 16-bit audio \cite{oord2018parallel}. Training was run for a total of 1,900,000 iterations, and took approximately 200 hours on a GPU. At inference time, a beam-search algorithm \cite{huang2018timbretron} was used to remove spurious impulse events generated by the probabilistic sampling method.


\subsection{Dataset}
To include a diverse range of sounds, a dataset was constructed from three source datasets. The datasets used were the ESC-50 dataset \cite{piczak2015dataset}, a labeled collection of environmental audio recordings, the Freesound Loops dataset \cite{ramires2020}, a collection of short musical clips including electronic and acoustics instrument sounds,  as well as speech sounds from the LJ-speech dataset \cite{ljspeech17}. Samples were randomly removed from the LJ-speech and Freesound loops datasets, so the different classes of sounds were evenly represented in the combined dataset. In total, the dataset used for training contained approximately 4 hours of audio.

\section{Hybrid TSM Method}
\label{sec:tsm}
\begin{figure}[t!]
\centering
\includegraphics[width=\columnwidth]{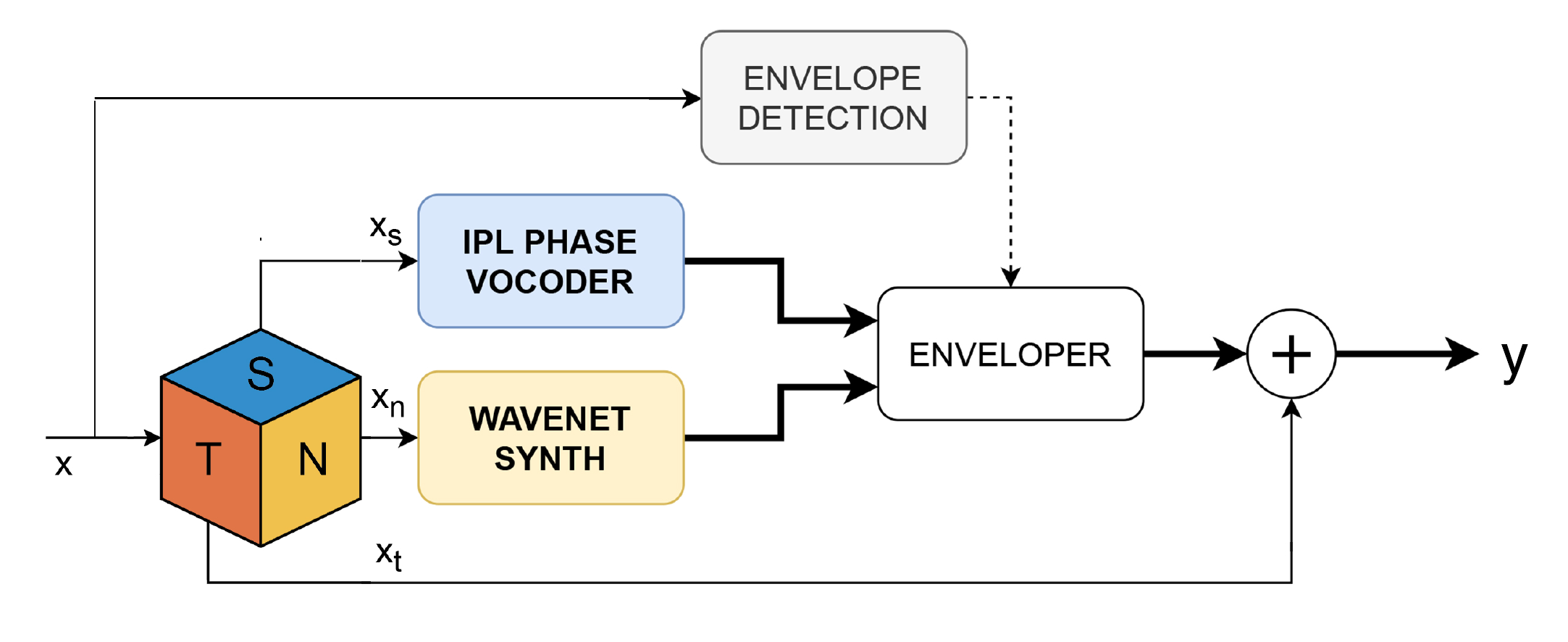}
\caption{Block diagram of the TSM pipeline, for a single-channel signal. A thicker line indicates a time-stretched signal.}
\label{fig:blockdiagram}
\end{figure}

The proposed model combines the use of traditional digital signal processing techniques and the Wavenet synthesizer to improve the time-scale modification performance for extreme time-stretching factor, and it is designed and trained to provide high quality stretching for environmental sounds. The TSM pipeline of the proposed method described in this section is visualized in Fig.~\ref{fig:blockdiagram}. 

\subsection{Processing the individual STN components}

The input audio signal is decomposed as described in Sec.~\ref{sec:STN}. Transition regions for the STN masks are defined by $\beta_\textrm{U}$ = [0.8 0.85] and $\beta_\textrm{L}$ = [0.7 0.75], where the two elements refer to the different transition region in each stage of the decomposition \cite{fierro2022enhanced}.

The sines are processed via phase vocoder with identity phase locking (IPL) \cite{laroche1999improved}, as used in \cite{damskagg2017audio}. Transients are preserved and relocated onto the new time axis \cite{nagel2009novel} according to detected peaks, which are isolated due to the nature of the STN decomposition. The noise component is neurally resynthesized at the desired TSM factor, as described in Sec.~\ref{sec:wavenet}.

\subsection{Post-processing}

Before mixing the three components, the time-stretched sines and noise are processed through an envelope (see Fig.~\ref{fig:blockdiagram}), which reshapes them according to the envelope of the original signal to compensate for the pre-echo effect that is typical of spectrogram-based TSM methods \cite{driedger2016review}. Finally, the three components are added and the time-stretched output is obtained, as also illustrated in Fig.~\ref{fig:blockdiagram}.

\begin{figure}[t!]
\centering
\includegraphics[width=\columnwidth]{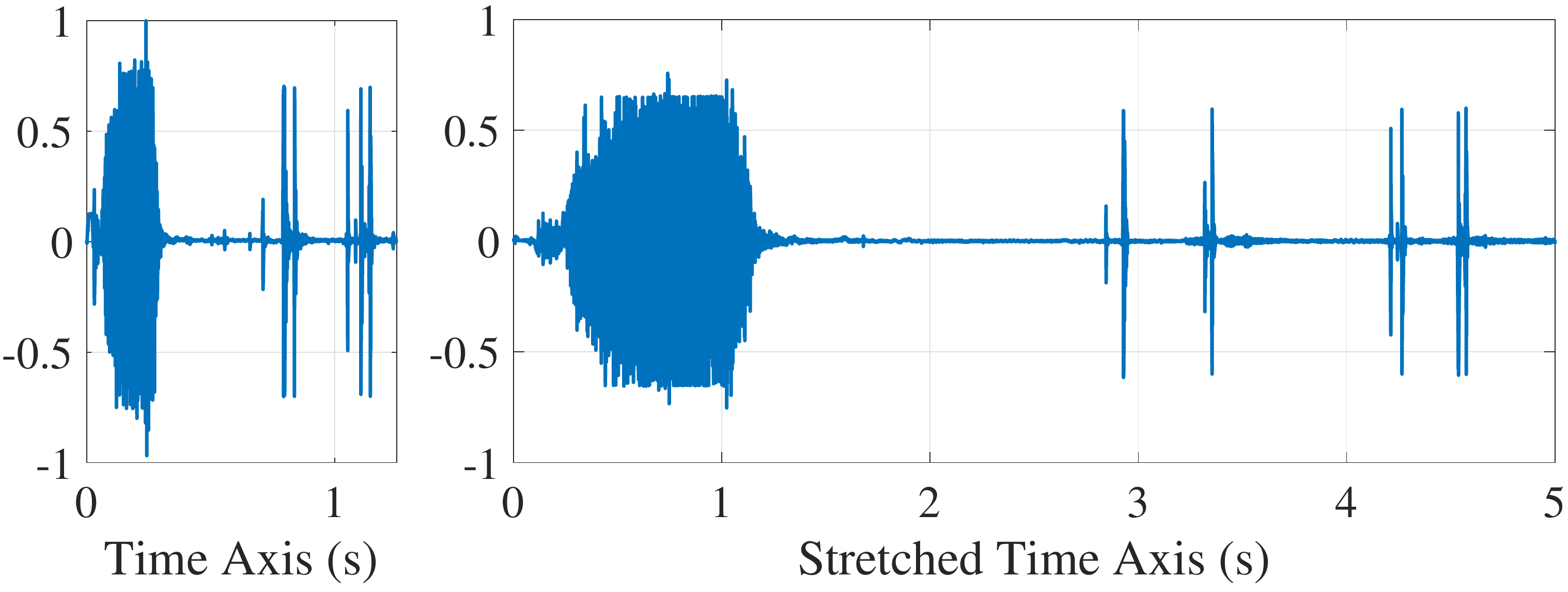}
\caption {(Left) \textit{Soda} sound and (right) its four-times stretched version processed via the proposed model.}
\label{fig:TSM}
\end{figure}

A time-stretching output example is shown in Fig.~\ref{fig:TSM} for the \textit{Soda} sound detailed in Table \ref{tab:samples} and a TSM factor $\alpha$ = 4. The ``hiss'' of the can opening is correctly resynthesized by the network, together with the noisy part of the later clicks which are imposed over the preserved transients. The STN decomposition is particularly helpful in this situation, as the merging of unaltered transients and the neurally-synthesized noise helps generate realistic sounds that retain the punch of percussive events while avoiding common artifacts, such as loss of presence, metallic tones, and phasiness.

\subsection{Stereo compatibility}

The process described so far can be individually applied to the two channels of a stereo sound, as STN decomposition is transparent with respect to the stereo field. When a sound is decomposed into its individual components and then recomposed, the stereo balance and phase coherence between the left and the right channel, computed according to \cite{roberts2018stereo} are preserved. This allows for each stereo channel to be processed independently.

\begin{table}[t!]
\centering
\caption{Audio samples used in the listening test.}
\begin{tabular}{ll}
\hline
\textbf{Name} & \textbf{Description} \\ \hline
Fireworks        & Two fireworks exploding in the air \\ 
Soda             & Hiss and click sound from a can opening  \\ 
PingPong         & A clip from an amateur ping pong game \\ 
Saw              & Handsaw sawing through wood \\ 
Sneeze           & A person sneezing two times \\ 
Rooster          & A rooster crowing in the morning \\ \hline
\end{tabular}
\label{tab:samples}
\end{table}

\begin{figure*}[t!]
\centering
\begin{subfigure}[t!]{\columnwidth}
\includegraphics[width=\columnwidth]{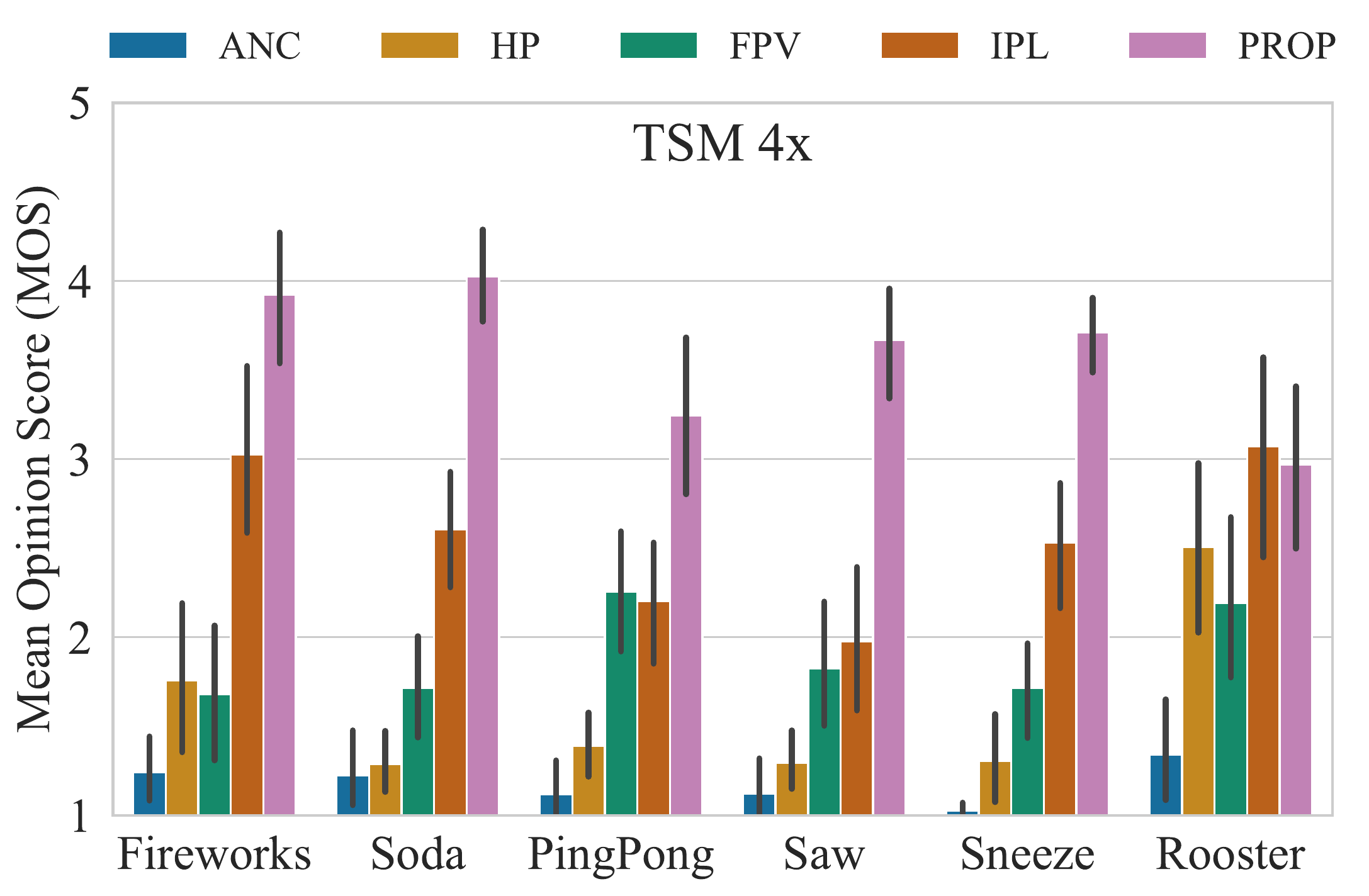}
\caption{}
\label{fig:barplot4}
\end{subfigure}
\begin{subfigure}[t!]{\columnwidth}
\includegraphics[width=\columnwidth]{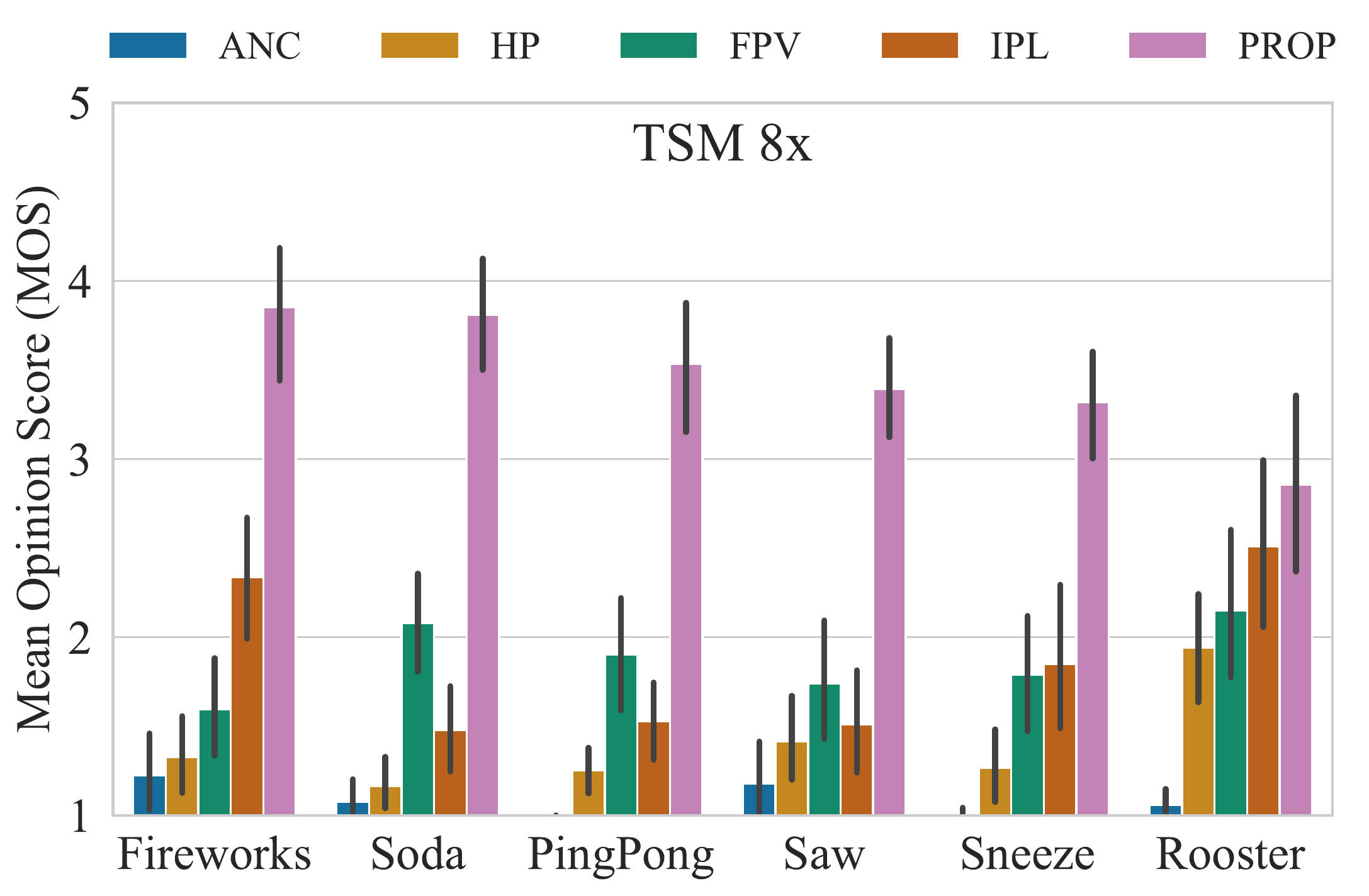}
\caption{}
\label{fig:barplot8}
\end{subfigure}
\caption{Barplots relative to the results of the subjective listening test, showing the average mean opinion scores and 95\% confidence intervals of the data for TSM factors (a) $\alpha$ = 4 and (b) $\alpha$ = 8. The proposed method outperforms the opposing algorithms in both cases for most of the samples under test.}
\label{fig:barplots}
\end{figure*}

\section{Evaluation}
\label{sec:evaluation}

The proposed method has been validated and compared with previous TSM methods in a listening test, which is reported in this section.

\subsection{Listening test design}

A formal blind listening test was conducted on a selection of 14 experienced listeners, all of which reported previous experience in test design and no hearing impairment or other relevant medical condition. 
The test software, a customized version of WebMushra \cite{schoeffler2018webmushra}, was run on a desktop machine running MacOS 10.14.6, using a single pair of Sennheiser HD 650 headphones, inside a sound--proof listening booth at the Aalto Acoustics Lab, Espoo, Finland. 
A set of six audio samples, listed in Table \ref{tab:samples}, were selected. As the test involves extreme stretching factors, of up to 8 times, a very short duration for each sample (approximately 2\,s) was necessary to ensure that even the longest time-stretched sounds remained below 18\,s long. 

The proposed method (PROP) was evaluated against the fuzzy phase vocoder (FPV) \cite{damskagg2017audio}, harmonic-percussive TSM (HP) \cite{driedger2013improving}, the two-step phase vocoder with identity phase locking \cite{laroche1999improved} (IPL), and WSOLA \cite{verhelst1993overlap}, which was provided as the low-quality anchor (ANC). The last three algorithms are included in the TSM Toolbox \cite{driedger2014tsm}. All the processed sounds were loudness normalized according to ITU-BS.1770 recommendation \cite{ITU} to prevent loudness differences from affecting the grades.
In each trial of the test, subjects were presented with one of the audio excerpts, referred to as \textit{reference}, and were asked to blindly rate the quality of the time-scaling operation. The original reference was not included among the samples to be rated. Subjects were instructed to rate each sample on a scale from 0 to 100, with no obligation of using the full scale, as the concept of perfect TSM is undefined and it was anticipated that none of the samples would be perceived to be an ideal processing.

The test was divided in two parts of twelve trials (six trials repeated twice for consistency) each, the first presenting a TSM factor of 4 and the second a TSM factor of 8, for a total of 24 trials and 72 audio samples under investigation.
Listeners were allowed a short training session before starting the actual test to get acquainted with the interface, the keyboard shortcuts, and the task itself. The results of the training session were not included in the statistical analysis. Prior to the test, familiarity of the subjects with the concepts of time-scale modification and transient smearing was assessed.

\subsection{Results}

Mean Opinion Scores (MOS) were computed from the ratings given by the subjects to estimate the quality of the time-stretching for the methods under test. Bar plots displaying the mean and 95\% confidence intervals of the data for are shown, in Fig.~\ref{fig:barplots}. The proposed method clearly outperforms the other algorithms under test under both time-stretching conditions for all samples but \textit{Rooster}. The majority of the test participants commented on this audio excerpt that the noisy nature and the lack of sharp transients of the reference made it hard to generate an expectation for a time-stretched version.

\section{Conclusion}
\label{sec:conclusion}

In this paper, we present a novel algorithm for extreme TSM, which takes advantage of the decomposition of sound into sines, transients, and noise to individually resynthesize the noise component using a deep neural network. Sines and transients are time-stretched using established signal processing techniques which also benefit from the decomposition. The results of a subjective listening test suggest that the proposed algorithm performs significantly better than previous TSM methods for real-world environmental recordings, when a large time-stretching factor is used. 
Future work involves redesigning the neural synthesis for standard time-stretching factors, involving a larger dataset of more relevant sounds. 


\bibliographystyle{IEEEbib}
\bibliography{refs}

\begin{thebibliography}{10}

\bibitem{moulines1995non}
E.~Moulines and J.~Laroche,
\newblock ``Non-parametric techniques for pitch-scale and time-scale
  modification of speech,''
\newblock {\em Speech Commun.}, vol. 16, no. 2, pp. 175--205, 1995.

\bibitem{dutilleux2011time}
P~Dutilleux, G~De~Poli, A~von~dem Knesebeck, and U~Z{\"o}lzer,
\newblock ``Time-segment processing (chapter 6),''
\newblock {\em DAFX: Digital Audio Effects, Second Edition; Z{\"o}lzer, U.,
  Ed}, pp. 185--217, 2011.

\bibitem{driedger2016review}
J.~Driedger and M.~M{\"u}ller,
\newblock ``A review of time-scale modification of music signals,''
\newblock {\em Appl. Sci.}, vol. 6, no. 2, pp. 57, 2016.

\bibitem{verhelst1993overlap}
W.~Verhelst and M.~Roelands,
\newblock ``An overlap-add technique based on waveform similarity ({WSOLA}) for
  high quality time-scale modification of speech,''
\newblock in {\em Proc. IEEE Int. Conf. Acoust. Speech Signal Process.
  (ICASSP)}, 1993, vol.~2, pp. 554--557.

\bibitem{donnellan2003speech}
O.~Donnellan, E.~Jung, and E.~Coyle,
\newblock ``Speech-adaptive time-scale modification for computer assisted
  language-learning,''
\newblock in {\em Proc. 3rd IEEE Int. Conf. Advanced Learning Technologies},
  Athens, Greece, 2003, pp. 165--169.

\bibitem{cohen2022speech}
E.~Cohen, F.~Kreuk, and J.~Keshet,
\newblock ``Speech time-scale modification with {GANs},''
\newblock {\em IEEE Signal Process. Lett.}, vol. 29, pp. 1067--1071, 2022.

\bibitem{cliff2000hang}
D.~Cliff,
\newblock ``Hang the {DJ}: Automatic sequencing and seamless mixing of
  dance-music tracks,''
\newblock {\em HP Laboratories Technical Report}, vol. 104, 2000.

\bibitem{nam2018deep}
J.~Nam, K.~Choi, J.~Lee, et~al.,
\newblock ``Deep learning for audio-based music classification and tagging:
  Teaching computers to distinguish rock from {Bach},''
\newblock {\em IEEE Signal Process. Mag.}, vol. 36, no. 1, pp. 41--51, Jan.
  2019.

\bibitem{damskagg2017audio}
E-P. Damsk{\"a}gg and V.~V{\"a}lim{\"a}ki,
\newblock ``Audio time stretching using fuzzy classification of spectral
  bins,''
\newblock {\em Appl. Sci.}, vol. 7, no. 12, pp. 1293, Dec. 2017.

\bibitem{roberts2021deep}
T.~Roberts, A.~Nicolson, and K.~K. Paliwal,
\newblock ``Deep learning-based single-ended quality prediction for time-scale
  modified audio,''
\newblock {\em J. Audio Eng. Soc.}, vol. 69, no. 9, pp. 644--655, Sept. 2021.

\bibitem{laroche1997phase}
J.~Laroche and M.~Dolson,
\newblock ``Phase-vocoder: About this phasiness business,''
\newblock in {\em Proc. IEEE Workshop Appl. Signal Processing to Audio and
  Acoustics (WASPAA)}, New Paltz, NY, Oct. 1997.

\bibitem{moinet2013slowdio}
A.~Moinet,
\newblock {\em Slowdio: Audio Time-Scaling for Slow Motion Sports Videos},
\newblock Ph.D. thesis, University of Mons, Mons, Belgium, 2013.

\bibitem{Valimaki2018}
V.~V{\"a}lim{\"a}ki, J.~R{\"a}m{\"o}, and F.~Esqueda,
\newblock ``Creating endless sounds,''
\newblock in {\em Proc. 21st Int. Conf. Digital Audio Effects (DAFx)}, Aveiro,
  Portugal, Sep. 2018, pp. 32--39.

\bibitem{malloy2022timbral}
C.~Malloy,
\newblock ``Timbral effects: {T}he {Paulstretch} audio time-stretching
  algorithm,''
\newblock {\em J. Acous. Soc. Am.}, vol. 151, no. 4, pp. A158--A158, 2022.

\bibitem{fierro2022enhanced}
L.~Fierro and V.~V{\"a}lim{\"a}ki,
\newblock ``Enhanced fuzzy decomposition of sound into sines, transients, and
  noise,''
\newblock {\em arXiv preprint arXiv:2210.14041}, Oct. 2022.

\bibitem{oord2016wavenet}
A.~van~den Oord, S.~Dieleman, H.~Zen, K.~Simonyan, et~al.,
\newblock ``{WaveNet: A} generative model for raw audio,''
\newblock {\em arXiv preprint arXiv:1609.03499}, Sept. 2016.

\bibitem{huang2018timbretron}
S.~Huang, Q.~Li, C.~Anil, et~al.,
\newblock ``{TimbreTron}: {A WaveNet(CycleGAN(CQT}(audio))) pipeline for
  musical timbre transfer,''
\newblock {\em arXiv preprint arXiv:1811.09620}, May 2019.

\bibitem{fitzgerald2010harmonic}
D.~Fitzgerald,
\newblock ``Harmonic/percussive separation using median filtering,''
\newblock in {\em Proc. Int. Conf. Digital Audio Effects (DAFx)}, Graz,
  Austria, 2010.

\bibitem{brown1991calculation}
J.~C. Brown,
\newblock ``Calculation of a constant {Q} spectral transform,''
\newblock {\em J. Acoust. Soc. Am.}, vol. 89, no. 1, pp. 425--434, 1991.

\bibitem{huzaifah2017comparison}
M.~Huzaifah,
\newblock ``Comparison of time-frequency representations for environmental
  sound classification using convolutional neural networks,''
\newblock {\em arXiv preprint arXiv:1706.07156}, 2017.

\bibitem{oord2018parallel}
A.~van~den Oord, Y.~Li, I.~Babuschkin, K.~Simonyan, et~al.,
\newblock ``Parallel {WaveNet}: Fast high-fidelity speech synthesis,''
\newblock in {\em Proc. Int. Conf. Machine Learning}, July 2018, pp.
  3918--3926.

\bibitem{piczak2015dataset}
K.~J. Piczak,
\newblock ``{ESC}: {Dataset} for environmental sound classification,''
\newblock in {\em Proc. 23rd Annual ACM Conf. Multimedia}, Oct. 2015, pp.
  1015--1018.

\bibitem{ramires2020}
A.~Ramires, F.~Font, D.~Bogdanov, et~al.,
\newblock ``The {Freesound} loop dataset and annotation tool,''
\newblock in {\em Proc. 21st Int. Conf. Music Information Retrieval (ISMIR)},
  2020.

\bibitem{ljspeech17}
K.~Ito and L.~Johnson,
\newblock ``The {LJ} speech dataset,''
  \url{https://keithito.com/LJ-Speech-Dataset/}, 2017.

\bibitem{laroche1999improved}
J.~Laroche and M.~Dolson,
\newblock ``Improved phase vocoder time-scale modification of audio,''
\newblock {\em IEEE Trans. Speech and Audio Process.}, vol. 7, no. 3, pp.
  323--332, 1999.

\bibitem{nagel2009novel}
F.~Nagel and A.~Walther,
\newblock ``A novel transient handling scheme for time stretching algorithms,''
\newblock in {\em Proc. Audio Eng. Soc. 127th Conv.}, 2009.

\bibitem{roberts2018stereo}
T.~Roberts and K.~K. Paliwal,
\newblock ``Stereo time-scale modification using sum and difference
  transformation,''
\newblock in {\em Proc. 12th Int. Conf. Signal Process. Comm. Syst. (ICSPCS)},
  Dec. 2018, pp. 1--5.

\bibitem{schoeffler2018webmushra}
M.~Schoeffler, S.~Bartoschek, F-R. St{\"o}ter, et~al.,
\newblock ``{WebMUSHRA}---{A} comprehensive framework for web-based listening
  tests,''
\newblock {\em J. Open Research Software}, vol. 6, no. 1, 2018.

\bibitem{driedger2013improving}
J.~Driedger, M.~M{\"u}ller, and S.~Ewert,
\newblock ``Improving time-scale modification of music signals using
  harmonic-percussive separation,''
\newblock {\em IEEE Signal Process. Lett.}, vol. 21, no. 1, pp. 105--109, 2013.

\bibitem{driedger2014tsm}
J.~Driedger and M.~M{\"u}ller,
\newblock ``{TSM Toolbox: MATLAB} implementations of time-scale modification
  algorithms,''
\newblock in {\em Proc. Int. Conf. Digital Audio Effects (DAFx)}, 2014, pp.
  249--256.

\bibitem{ITU}
{ITU-R BS.1770-4},
\newblock ``{Algorithms to measure audio programme loudness and true-peak audio
  level},''
\newblock standard, ITU/Radiocommunications, Oct. 2015.

\end{thebibliography}

\end{document}